\documentclass[aps,showpacs,twocolumn]{revtex4}

\begin{document}

\title{Comment on ``Dynamical mass generation in strongly coupled
quantum electrodynamics with weak magnetic fields''}

\author{S.-Y. Wang}
%\email{sywang@mail.tku.edu.tw}

\affiliation{Department of Physics, Tamkang University, Tamsui,
Taipei 25137, Taiwan}

\date{June 20, 2006}

\begin{abstract}
In a recent article, Ayala \emph{et al}.\ [Phys.\ Rev.\ D
\textbf{73}, 105009 (2006)] have studied the dynamical generation of
masses for fundamental fermions in strongly coupled QED in the
presence of weak magnetic fields. We argue that the results and
conclusions of the article are not reliable as they are subject to
gauge dependent artifacts resulting from an inconsistent truncation
of the Schwinger-Dyson equations.
\end{abstract}


\maketitle

Ayala \emph{et al}.~\cite{Ayala:2006sv} have recently studied
dynamical chiral symmetry breaking in strongly coupled QED in the
presence of a weak, external magnetic field. Not long ago, a similar
problem in QED in a strong, external magnetic field was critically
studied by the present author~\cite{Leung:2005xz,Leung:2005yq}. We
carefully examined the consistency and gauge independence of the
bare vertex approximation (BVA) that has been extensively used in
truncating the Schwinger-Dyson (SD) equations to calculate the
dynamically generated fermion mass in the lowest Landau level
approximation (LLLA). Specifically, we showed that the BVA, in which
the vertex corrections are ignored, is a consistent truncation of
the SD equations in the LLLA. The dynamical fermion mass, obtained
as the solution of the truncated SD equations evaluated on the
fermion mass shell, is manifestly gauge independent. Based on the
gauge independent analysis presented in
Refs.~\cite{Leung:2005xz,Leung:2005yq}, we argue that the results
and conclusions of Ref.~\cite{Ayala:2006sv} are not reliable as they
are subject to gauge dependent artifacts resulting from an
inconsistent truncation of the SD equations.

The important lessons learned from the gauge independent analysis
presented in Refs.~\cite{Leung:2005xz,Leung:2005yq} are as follows.
(i) A consistent truncation of the SD equations is the truncation
which respects the Ward-Takahashi (WT) identity satisfied by the
truncated vertex and inverse fermion propagator. (ii) In a
consistent truncation of the SD equations the WT identity is a
necessary (but not sufficient) condition for establishing the gauge
independence of physical observables calculated therein. Therefore,
if one begins with an \emph{inconsistent} truncation of the SD
equations, in which the WT identity is not fulfilled, then it is
impossible to obtain gauge independent results for physical
observables (e.g., the dynamical fermion mass for the problem at
hand). As a result, thorough investigations of the gauge
(in)dependence of the calculated physical observables constitute an
indispensable part of a \emph{consistent} or a \emph{reliable}
analysis of the SD equations.

In Ref.~\cite{Ayala:2006sv} the BVA in quenched QED (or the
so-called rainbow approximation), in which the vertex corrections as
well as the vacuum polarization are ignored, was used to truncate
the SD equations \emph{beyond} the LLLA. Such an approximation
however is \emph{not} a consistent truncation of the SD equations as
it has been shown in Ref.~\cite{Leung:2005yq} that the WT identity
in the BVA can be satisfied only \emph{within} the LLLA (with a
momentum independent fermion self-energy). This is a general
statement that is valid regardless of the external magnetic field
strength or the gauge fixing. To put it another way, in order to go
beyond the LLLA (in either a strong or a weak magnetic field) one
has to use a truncation scheme of the SD equations that consistently
accounts for \emph{vertex corrections} (as well as the momentum
dependence of the fermion self-energy). Furthermore, the assertion
in Ref.~\cite{Ayala:2006sv} regarding the reliability of the WT
identity in the BVA beyond the LLLA in the Landau gauge for small
values of fermion momenta is invalid and misleading. While it is the
case in the LLLA as can be seen clearly in Fig.~1, in which the mass
function in the LLLA is depicted in the Landau gauge, it is
certainly \emph{not} the case when going beyond the LLLA as we have
explained above. In fact, Fig.~1 is simply a manifestation of the
reliability of a momentum independent fermion self-energy and,
consequently, of the WT identity in the BVA \emph{within} the LLLA
(see Ref.~\cite{Leung:2005yq} for detailed discussions). Contrary to
what the authors of Ref.~\cite{Ayala:2006sv} have claimed, Fig.~1
unfortunately reveals nothing about the reliability of the WT
identity in the BVA \emph{beyond} the LLLA.

A similar approximation with the BVA in unquenched QED (or the
so-called improved rainbow approximation) has been used in
Ref.~\cite{Kuznetsov:2002zq} to study \emph{higher} Landau level
corrections to the dynamical fermion mass generated in a strong
magnetic field. Instead of a verification of the lowest Landau level
(LLL) dominance in a strong magnetic field, the result of
Ref.~\cite{Kuznetsov:2002zq} is found to be \emph{qualitatively}
different from those obtained in the BVA within the LLLA. We have
argued~\cite{Leung:2005yq} that in the specific gauge used in
Ref.~\cite{Kuznetsov:2002zq} the unphysical, gauge dependent
contributions from higher Landau levels are so large that they
become dominant over the physical, gauge independent contribution
from the lowest Landau level and lead to the authors' incorrect
conclusions. Clearly, the particular choice of gauge fixing used in
Ref.~\cite{Kuznetsov:2002zq} not only is ad hoc and unnecessary, but
also leaves the issue of gauge (in)dependence unaddressed. We have
also pointed out~\cite{Leung:2005yq} that the LLL dominance in a
strong magnetic field (or, equivalently, lack thereof in a weak
magnetic field for the problem at hand) should be understood in the
context of a consistent truncation scheme: In a strong magnetic
field higher Landau level contributions to the dynamical fermion
mass obtained in a consistent truncation of the SD equations (which
accounts for vertex corrections) are subleading when compared to the
LLL one obtained in the consistent BVA truncation (which ignores
vertex corrections). The inconsistency of the BVA beyond the LLLA in
QED in the presence of an external magnetic field is \emph{not}
unexpected. This is because when going beyond the LLLA there is no
longer an effective dimensional reduction in the dynamics of fermion
pairing (see Ref.~\cite{Leung:2005yq} and references therein). The
situation is then similar to that in ordinary QED, in which it is
well known that the BVA is \emph{not} a consistent truncation of the
SD equations. As a consequence, in the BVA \emph{beyond} the LLLA
the results of Ref.~\cite{Ayala:2006sv} obtained in the Landau gauge
are inevitably gauge dependent artifacts. This in turn calls into
serious question about the reliability of the subsequent
conclusions.

Incidentally, on the technical side, the authors of
Ref.~\cite{Ayala:2006sv} have largely missed several important
issues regarding the properties of the Ritus functions (a convenient
formalism in the studies of QED in the presence of a constant
external magnetic field) that have been overlooked in the literature
and clarified only recently in Ref.~\cite{Leung:2005yq}. Hence, the
calculation in Ref.~\cite{Ayala:2006sv} will require further
investigations.

We conclude by emphasizing that \emph{in gauge theories an analysis
of the nonperturbative SD equations should not and cannot be
considered reliable or complete unless the gauge independence of
physical observables calculated therein is unambiguously
demonstrated}. When utilizing the SD equations in gauge theories, it
is therefore mandatory to begin in the first place with a consistent
truncation or, at least, a reliable (but not necessarily consistent)
truncation that yields a controlled gauge dependence with the
explicit gauge dependent terms appearing at higher order than the
truncation order~\cite{endnote}. Otherwise, it should be kept in
mind as well as stated explicitly that the obtained results are
inevitably gauge dependent and without further quantification of the
gauge dependent artifacts no reliable conclusions may be drawn
therefrom.

This work was supported in part by the National Science Council of
Taiwan under grant NSC-95-2112-M-032-010.


\begin{thebibliography}{9}

%\cite{Ayala:2006sv}
\bibitem{Ayala:2006sv}
A. Ayala, A. Bashir, A. Raya, and E. Rojas,
%``Dynamical mass generation in strongly coupled quantum electrodynamics with
%weak magnetic fields,''
Phys.\ Rev.\ D \textbf{73}, 105009 (2006).
%[arXiv:hep-ph/0602209].
%%CITATION = HEP-PH 0602209;%%

%\cite{Leung:2005xz}
\bibitem{Leung:2005xz}
C. N. Leung and S.-Y. Wang,
%``Gauge independence and chiral symmetry breaking in a strong magnetic
%field,''
hep-ph/0503298 (Ann. Phys., in press).
%%CITATION = HEP-PH 0503298;%%

%\cite{Leung:2005yq}
\bibitem{Leung:2005yq}
C. N. Leung and S.-Y. Wang,
%``Gauge independent approach to chiral symmetry breaking in a strong magnetic
%field,''
Nucl. Phys. \textbf{B747}, 266 (2006).
%[arXiv:hep-ph/0510066].
%%CITATION = HEP-PH 0510066;%%

%\cite{Kuznetsov:2002zq}
\bibitem{Kuznetsov:2002zq}
A.~V.~Kuznetsov and N.~V.~Mikheev,
%``Electron mass operator in a strong magnetic field and dynamical chiral
%symmetry breaking,''
Phys. Rev. Lett. \textbf{89}, 011601 (2002).
%[arXiv:hep-ph/0204201].
%%CITATION = HEP-PH 0204201;%%

\bibitem{endnote}
For related discussions on reliable truncations in (high
temperature) gauge theories, see, for instance,
%
%\cite{Arrizabalaga:2002hn}
%\bibitem{Arrizabalaga:2002hn}
A.~Arrizabalaga and J.~Smit,
%``Gauge-fixing dependence of Phi-derivable approximations,''
Phys. Rev. D \textbf{66}, 065014 (2002);
%[arXiv:hep-ph/0207044].
%%CITATION = HEP-PH 0207044;%%
%
%\cite{Aarts:2002tn}
%\bibitem{Aarts:2002tn}
G.~Aarts and J.~M.~Martinez Resco,
%``Ward identity and electrical conductivity in hot QED,''
JHEP \textbf{0211}, 022 (2002);
%[arXiv:hep-ph/0209048].
%%CITATION = HEP-PH 0209048;%%
%
%\cite{Mottola:2003vx}
%\bibitem{Mottola:2003vx}
E.~Mottola,
%``Gauge invariance in 2PI effective actions,''
in \emph{Proceedings of the Strong and Electroweak Matter 2002
Meeting} (World Scientific, Singapore, 2003) [hep-ph/0304279];
%%CITATION = HEP-PH 0304279;%%
%
%\cite{Boyanovsky:2002te}
%\bibitem{Boyanovsky:2002te}
D.~Boyanovsky, H.~J.~de Vega, and S.-Y.~Wang,
%``Dynamical renormalization group approach to transport in ultrarelativistic
%plasmas: The electrical conductivity in high temperature QED,''
Phys.\ Rev.\ D \textbf{67}, 065022 (2003);
%[arXiv:hep-ph/0212107].
%%CITATION = HEP-PH 0212107;%%
%
%\cite{Wang:2004tg}
%\bibitem{Wang:2004tg}
S.-Y.~Wang,
%``Gauge dependence of the fermion quasiparticle poles in hot gauge
%theories,''
Phys.\ Rev.\ D \textbf{70}, 065011 (2004);
%[arXiv:hep-ph/0406002].
%%CITATION = HEP-PH 0406002;%%
%
%\cite{Carrington:2003ut}
%\bibitem{Carrington:2003ut}
M.~E.~Carrington, G.~Kunstatter, and H.~Zaraket,
%``2PI effective action and gauge invariance problems,''
Eur. Phys. J. \textbf{C42}, 253 (2005).
%%CITATION = HEP-PH 0309084;%%

\end{thebibliography}
\end{document}